**TITLE: Process reveals structure: How a network is traversed mediates expectations about its architecture**

AUTHORS: Elisabeth A. Karuza[1*], Ari E. Kahn[2,3], Sharon L. Thompson-Schill[1,4], & Danielle S. Bassett[3,5]

AFFILIATIONS:
[1]Department of Psychology, University of Pennsylvania, Philadelphia, PA 19104 USA
[2]Department of Neuroscience, University of Pennsylvania, Philadelphia, PA 19104 USA
[3]Department of Bioengineering, University of Pennsylvania, Philadelphia, PA 19104 USA
[4]Department of Neurology, University of Pennsylvania, Philadelphia, PA 19104 USA
[5]Department of Electrical and Systems Engineering, University of Pennsylvania, Philadelphia, PA 19104 USA

*CORRESPONDING AUTHOR:
Elisabeth A. Karuza
ekaruza@sas.upenn.edu

**Abstract**

Network science has emerged as a powerful tool through which we can study the higher-order architectural properties of the world around us. How human learners exploit this information remains an essential question. Here, we focus on the temporal constraints that govern such a process. Participants viewed a continuous sequence of images generated by three distinct walks on a modular network. Walks varied along two critical dimensions: their predictability and the density with which they sampled from communities of images. Learners exposed to walks that richly sampled from each community exhibited a sharp increase in processing time upon entry into a new community. This effect was eliminated in a highly regular walk that sampled exhaustively from images in short, successive cycles (i.e., that increasingly minimized uncertainty about the nature of upcoming stimuli). These results demonstrate that temporal organization plays an essential role in how robustly knowledge of network architecture is acquired.

## Introduction

From classic computational models[1,2] to the exploration of human brain dynamics[3,4], network science and cognitive science continually inform and advance one another. In network science, one essential question concerns how real-word external systems, from social groups[5] to natural language[6], are structured. In cognitive science, network-based frameworks have been applied to uncover the internal architecture that gives rise to human behavior [7]. Here, we focus on a novel question that lies firmly at the intersection of network science and cognitive science, with far-reaching implications for the study of human learning: What constrains the extraction of higher-order properties from an external complex system? We address a persistent gap in the convergence of network and cognitive science by examining whether learning of a graph structure is influenced by its temporal organization, or the order in which its edges are traversed. First, we show that an implicit, trial-by-trial measure of behavior reflects the network topology underlying a given stimulus environment, in this case a continuous series of visual events. Crucially, by systematically altering how the elements of a network unfold in time, we then demonstrate that human learners rely on the path through a network as they generate expectations about its structure.

Across domains, emerging evidence suggests that learners are sensitive to the network properties of their surroundings. Measures such as the clustering coefficient of a word (an index of how closely its phonological neighbors are related to one another) have been shown to predict how robustly that word is ultimately acquired[8]. Additional evidence indicates that the distance between objects in a network based on their physical features influences how early those objects are labeled by young children[9]. Outside of the language domain, adult learners have been shown to exploit higher-order temporal relations to segment continuous streams of images into visual events. In related prior work[10], participants were exposed to a sequence of images generated by a random walk on a network comprised of three groupings (communities) of densely interconnected nodes. Because each node had equivalent degree (number of edges incident to that node), a random walk on this network resulted in a continuous sequence of images with equivalent pairwise transition probabilities (TPs). In the statistical learning literature, peaks and dips in TPs constitute a powerful cue to event structure[11–13]; low TPs are considered an index of unpredictability, which in turn signals an event boundary. Nonetheless, even in the absence of this key source of information, learners were able to explicitly segment the stream into distinct communities after an extended training phase.

The afore-mentioned findings offer important insight into learners' sensitivity to complex patterns over and above local associations between elements in their environment (for a detailed discussion of this topic see[14]). Network-based approaches thus have great potential to answer fundamental questions about the scale at which learning operates. However, if this line of inquiry is to continue to flourish, it is also essential to probe the process through which higher order patterns are revealed to the learner. While learning is clearly influenced by the topological properties underlying sensory input, the present study addresses whether learning is additionally influenced by the order (walk) in which the edges of a graph are traversed. Phrased another way, how does the distribution of information in time affect learning of an underlying network architecture?

A largely untapped method of examining the influence of walk structure is to continuously monitor learners' processing speed as the elements of a graph-based sequence unfold. To the extent it varies by walk structure, an increase in reaction time (RT) associated with the unexpected (i.e., a surprisal effect) would serve as compelling evidence that learners are indeed sensitive to the process through which higher order patterns are revealed to them (this premise has been richly explored in the language domain[15]; for a discussion of surprisal in the context of syntactic comprehension see[16]). Though not necessarily framed in terms of expectation generation, related reaction time techniques have also been used to examine the acquisition of complex motor sequences[17].

In the present study, we exposed adult learners to sequences of fractal images generated by three distinct walks on an identical underlying graph: a random walk, a walk consisting of successive Eulerian paths, and a walk consisting of successive Hamiltonian paths. In contrast to a random walk, Eulerian and Hamiltonian paths are highly structured, sampling exhaustively from all edges and nodes in a network, respectively (Fig. 1). The influence of temporal organization was examined using a graph with dense community structure[10]; nodes on that graph each represented a distinct visual image and edges connecting pairs of nodes represented their co-occurrence over a period of continuous exposure. Random, Hamiltonian, and Eulerian walk types were selected to probe two dimensions we hypothesized to be of great importance to the learner: uncertainty and redundancy. Because of the highly structured nature of the Eulerian and Hamiltonian paths, uncertainty was reduced relative to the Random condition (particularly rapidly for the Hamiltonian paths, which involved presentations of nodes in series of only 15). As each path progressed, the probability that a given node or edge would be encountered increased steadily (i.e., if it was not previously encountered in that series). Thus, as the set of possible traversals through the network was systematically narrowed, uncertainty about upcoming stimuli also decreased. On the other hand, we hypothesized that redundancy within a sequence might also constitute an important cue to event structure. In contrast to Hamiltonian paths, Eulerian paths and random walks tend to stay inside a given community, resulting in prolonged and repeated exposure to the common connections between nodes within a cluster. Exposure to these common connections, when presented in close temporal proximity, might then serve as a crucial learning cue.

During exposure to each walk type, participants completed a task requiring them to indicate via button press whether or not each image was rotated away from its canonical orientation. We then examined RTs to determine whether learners were developing community-based expectations about upcoming stimuli. Specifically, we contrasted transition nodes, which represented entry into a new community of images, with pre-transition nodes, the images that directly preceded them. If learners were successfully developing higher-order expectations, then we should observe an increase in processing time for the transition nodes. Below, we detail evidence that expectation- based processing indeed operates at the community level, but we implicate a crucial role for redundancy in extracting the topological properties of a densely clustered network.

## Results

**Rotation Detection**

Participants generally excelled in distinguishing between the rotated and canonical fractal images during the exposure phase. Results indicate that cover task compliance was high across conditions (Fig. 2; mean $A'$ = 0.90, s.d. = 0.08; versus chance, $t(59) = 90.88$, $p < 0.001$).

**Model 1: Random *versus* Hamiltonian Paths**

When comparing RTs for the Random and Hamiltonian conditions (Model 1), we found significant main effects of Node Type ($\beta = 10.85$, $t = 6.28$, $p < 0.001$) and Trial ($\beta = -31.27$, $t = -7.72$, $p < 0.001$). Overall, RTs for pre-transition nodes were facilitated relative to transition nodes, and, as anticipated, participants generally sped up over the course of the experiment. Crucially, we observed a significant interaction between Node Type and Condition ($\beta = -8.53$, $t = -4.94$, $p < 0.001$), indicating that learners in the Random condition exhibited a significantly greater cross-community increase in processing time (referred to as a surprisal effect in Fig. 2) relative to learners in the Hamiltonian condition. This result is particularly impactful given that transition probabilities between pairs of nodes were equated in the Random condition. Finally, a significant interaction between Condition and Trial demonstrated that across both node types, learners were increasingly faster at processing elements in the Random condition ($\beta = 8.28$, $t = 2.05$, $p = 0.048$). Results for all main effects and their interactions are summarized in Table 1. Intriguingly, the Condition*Trial interaction was also maintained when examining all nodes in the network, as opposed to solely the boundary nodes ($\beta = 11.35$, $t = 2.94$, $p = 0.005$). Thus, despite the surprisal effects associated with transition nodes in the Random condition, overall RTs sped-up more quickly over time relative to the Hamiltonian condition.

Since Model 1 revealed a significant interaction between Node Type* Condition and Condition*Trial, a simple effects analysis was used to isolate the effects of Node Type and Trial separately for each condition. Results revealed that the Node Type*Condition interaction could be traced to a significant effect of Node Type exclusive to the Random condition ($\beta = 19.38$, $t = 6.78$, $p < 0.001$). No effect of Node type was observed for the Hamiltonian condition ($\beta = 2.33$, $t = 1.20$, $p = 0.236$). In contrast, the significant Condition*Trial interaction was not attributed to a null effect of Trial in the Hamiltonian condition ($\beta = -22.99$, $t = -4.14$, $p = 0.002$). Rather, the magnitude of the Trial effect was simply larger in the Random condition ($\beta = -39.55$, $t = -6.71$, $p < 0.001$).

**Model 2: Random *versus* Eulerian Paths**

Significant main effects of Node Type and Trial were maintained when examining the Random and Eulerian conditions. However, neither the Node Type by Condition interaction ($\beta = -0.61$, $t = -0.32$, $p > 0.250$) nor the Condition by Trial interaction ($\beta = 5.55$, $t = 1.44$, $p = 0.157$) were significant. Thus, neither the magnitude of the cross-community RT increase nor the overall speed-up of RTs in time differed between walk types that shared the property of redundancy. Results are summarized in Table 1. As we found no differences between the Eulerian and Random conditions, we did not perform a repetition priming analyses on Model 2 (see below).

**Repetition Priming**

Although we were interested in stimulus history effects as they relate to the network structures participants were learning, other types of stimulus history effects, such as item-specific repetition priming, could affect our observed measures. Processing times associated with a stimulus are known to decrease with recent exposure to that stimulus[18], which could explain a relative increase in RTs for transition relative to pre-transition nodes (as previously suggested[10]). To be clear, repeated instances of the same node within a short time frame might in fact serve as an important part of the learning mechanisms driving expectations about community structure. However, here we take a conservative approach in attempting to exclude low-level perceptual priming as the sole driver of pre-transition facilitation effects.

After accounting for repetition priming effects in the Random versus Hamiltonian comparison (**Materials and Methods**), each of the significant main effects (Node Type: $\beta = 6.26$, $t = 3.52$, $p < 0.001$ and Trial: $\beta = -31.39$, $t = -7.94$, $p < 0.001$), as well as the interactions between Node Type*Condition ($\beta = -4.40$, $t = -2.48$, $p = 0.014$) and Condition*Trial ($\beta = 8.33$, $t = 2.11$, $p = 0.041$), were maintained. Thus, though participants clearly were sensitive to repetition priming effects (i.e., *how many times each image was seen in the previous 10 trials*: β = –6.03, t = –3.00, p = 0.003; and *how many trials elapsed since each image was seen*: β = 7.71, t = 3.23, p = 0.002), any shared variance between these predictors and our interactions of interest was not sufficient to explain cross-community RT increases. After we confirmed that our significant interactions were maintained when accounting for priming, we verified via model comparison that this full model provided a superior fit to the data relative to a model that did not include main effects or interactions of Condition and Node Type (i.e., that only included both repetition priming predictors and Trial number). Indeed, a comparison of the log likelihood ratio between the two models was significant ($\chi2$ = 19.66, p = 0.003). These results indicate that it is essential to account for walk type in evaluating cross-community surprisal effects.

## Discussion

Implicit measures of processing speed during learning revealed that, even without variations in pairwise probabilistic information, learners exhibited a sharp increase in RT when transitioning out of a community of images. Because this effect was maintained when accounting for image-specific repetition priming, it suggests that learners successfully generated expectations reflective of the meso-level (community) structure underlying a sequence of visual events. Notably, evidence for community-based surprisal was limited to the Random and Eulerian conditions, both of which involved rich and repeated exposure to common connections within the same community. This property, which we term redundancy, represents a higher-order property than image-specific repetition priming. When the traversal of intra-community edges was sparsely distributed in time, as in the Hamiltonian condition, learners did not exhibit the same increase in RT upon entry into a new community. Moreover, though the Hamiltonian paths were rigidly structured, we suggest that the nature of this structure actually minimized uncertainty to the detriment of the learner (for a related discussion of ordering effects in a different sort of learning context, see[19]). While Eulerian paths also increasingly minimized uncertainty, they did so more gradually, operating over series of 30 edge traversals as opposed to 15 nodes. Thus, the redundancy inherent to the Eulerian paths was likely sufficient to override these comparatively subtle decreases.

Taken together, our results are powerful for a number of reasons. First, they indicate that, given proper temporal organization, learners develop remarkably high-level expectations about the nature of the upcoming signal. Compellingly, they do so even when pairwise transition probabilities between adjacent and non-adjacent elements are equated. The claim that prediction-based processes operate at a larger scale than local relationships accords with evidence from the sentence processing literature[15]. However, similar effects had not been systematically investigated, particularly from a graph theoretical perspective, outside the language domain. Indeed, the properties we have presently examined through the manipulation of walk structure have ties to a much wider literature on language processing and production[16,20,21]. Second, these findings shed light on a longstanding question surrounding the optimal learning paths in information networks[22]. A combination of computational modeling[23] and measures of semantic fluency[24] in humans indicate that retrieval processes operating on densely clustered semantic networks are optimized by algorithms that sample densely from those clusters before moving to another. Here, we have extended this work to the study of learning, offering compelling evidence that redundancy in walk structure spurs the acquisition of temporal communities in the context of event segmentation. Thus, at least in a dynamic sensory environment, the fixed graph topology underlying incoming stimuli does not itself appear sufficient for learning. Rather, the order in which that topology is revealed to the learner facilitates (or impedes) extraction of higher-order architectural properties.

## Materials and Methods

**Participants**

As verified by their assigned worker ID, 60 unique participants (20 per experimental condition) completed this study through Amazon Mechanical Turk, an on-line marketplace in which adult workers can perform behavioral experiments in exchange for financial compensation. Participants were paid at a rate of $0.08 per minute. To incentivize compliance with the task, they also received a completion bonus of $1.00 per phase of the experiment (exposure, segmentation, and odd-man out), and an additional $1.00 bonus if their cover task performance during the exposure phase exceeded 90% accuracy. Thus, total compensation for a participant who completed all phases of the experiment ranged from $7.00-$8.00. Participants communicated informed consent to engage in the study. Methods adhered to the guidelines and regulations of the Institutional Review Board (IRB) of the University of Pennsylvania. This committee (IRB at the University of Pennsylvania) also approved experimental protocols.

**Materials**

Stimuli consisted of 15 grayscale images generated via the Qbist filter in the GNU Image Manipulation program (v. 2.8.14; www.gimp.org). We elected to use unfamiliar, complex fractal images to minimize their nameability. Image-to-node assignment was randomized across participants.

*Exposure sequence.* In the Random condition, participants were exposed to a continuous sequence of 1400 images generated via a random walk on the graph shown in Fig. 1. This type of walk ensured that transition probabilities between pairs of nodes were equated, as each node in the graph was connected to precisely 4 other nodes. In the Hamiltonian condition, sequences were generated by concatenating a series of randomly selected forward and backward Hamiltonian paths, thereby ensuring that each of the 15 images was presented exactly once before a new path was initiated (1395 images total). The starting node of each path was a randomly selected node sharing an edge with the terminal node of the previous path. In the Eulerian condition, we concatenated a series of randomly selected forward and backward Eulerian paths (1395 images total), each of which visited every edge of the graph exactly once before initiating a new path. Note that although the Hamiltonian and Eulerian conditions consisted of a concatenation of shorter paths, the resulting sequence was entirely continuous.

**Procedure**

*Exposure phase.* Prior to initiating the exposure phase, participants were instructed as follows: "you will see a stream of abstract images flashed on the screen one at a time. Your job is to keep your eyes on the screen as the stream progresses. Over time, parts of the stream may become familiar to you. This part will take around 35 minutes." Participants were also informed that while they were viewing the images, they should indicate whether each image appeared in its canonical orientation (by pressing 1 on their keyboard) or whether it was rotated 90 degrees to the left (2 on their keyboard). Unknown to them, 15% of all images were rotated. To assist them on this task, participants first completed a study phase in which they viewed each image in its canonical orientation for 5 seconds. Next, they were asked to distinguish between a canonical and rotated image. Each trial was repeated until participants answered correctly. Furthermore, participants were instructed that they would hear a high-pitched tone if they responded incorrectly during the exposure phase and a low-pitch tone if they responded too slowly. Before beginning the exposure phase, participants had to pass a multiple-choice quiz that tested their knowledge of how that phase was structured. The full quiz was repeated until the participant achieved 100% accuracy on all questions. This step was intended to ensure high data quality, which is a potential concern for experiments conducted outside the laboratory[25]. To this end, participants were also informed that if they responded incorrectly (or failed to respond) to greater than 10 trials in a row, the experiment would terminate automatically. Participants were randomly assigned to one of two exposure lists per condition.

*Post-exposure measures.* After the exposure phase, participants completed two post-exposure tests: a segmentation task and an odd- man out judgment. These tests were intended to evaluate the explicit expression of graph structure knowledge. In the segmentation task, participants were asked only to press the space bar when they observed a natural breaking point in the stream of images (i.e., a new community). In the odd-man out judgment, participants were simultaneously presented with three images, two of which were drawn from the same community. They were asked to indicate which image "did not belong" with the others. Extensive detail and results from both tasks can be found in the **Supplementary Information Section 1**. As described there, data quality was much lower (loss of > 20% of participants), making it difficult to interpret the observed absence of learning effects. Lowered data quality was presumably due to a combination of fatigue effects and a lack of performance-based financial incentives for either post-exposure measure. The sum of these results highlights the importance of including financial incentives (beyond mere completion bonuses) for all phases of on-line experimentation [25].

**Analyses**

*Data exclusions.* Prior to implementing the mixed effects models described below, we eliminated all incorrect trials from the exposure phase (8.9% total data loss) followed by trials in which the image was rotated away from its canonical orientation (a further 11.0% loss). Next, the following pre-determined data trimming steps were performed. First, we removed implausible RTs (i.e., less than 100 ms; 2.7% loss). In accord with common approaches in reading time studies, we then removed outlier data points from each subject constituting > 3 standard deviations from their standardized mean processing times (1.5% data loss). We excluded all data from one subject from the Eulerian condition with an extremely anomalous RT pattern (i.e., the afore-mentioned data trimming techniques resulted in removal of 94.1% of that participant's data). All significant findings reported here hold without these data trimming techniques.

*Indexing cross-community surprisal.* We implemented a series of linear mixed effects models using the *lmer()* function (library lme4, v. 1.1-10) in R v. 3.2.2[26,27]. All data and corresponding analyses are available upon request. We focused on two types of community boundary nodes: transition nodes, which represented entry into a new community, and pre-transition nodes, which represented the node immediately prior to entry into a new community. If participants were indeed sensitive to the community structure of the network, then we should find a sharp increase in RT for the transition node relative to the pre-transition node. In rare cases involving a forward and backward traversal of the same edge (e.g., 5-6-5), we counted only the first of two transition nodes (in this case, 6).

Our analysis was split into two models, each of which compared one of the structured walks (either Hamiltonian or Eulerian) to the Random condition. We adopted a two model approach because the inclusion of all three walk types in the same model resulted in excessive multicollinearity between effects of interest, even when predictors were centered ($r$s were > 0.6).

RTs were regressed onto all main effects and interactions of Node Type (pre-transition versus transition), Condition (Random versus Hamiltonian or Random versus Eulerian) and Trial Number. All predictors were centered to reduce multicollinearity. Each model also included the fullest random effects structure that allowed the model to converge. For both Model 1 (Random versus Hamiltonian) and Model 2 (Random versus Eulerian), this corresponded to a random intercept for participant and by-participant random slopes for Trial, Node Type, and the interaction between the two.

*Accounting for repetition priming.* To the extent that recency of stimulus repetitions varied by walk type, we sought to measure priming effects and to distinguish these effects from any observed surprisal effects. We re-ran the model comparing the Random and Hamiltonian conditions, this time including two additional predictors intended to capture perceptual priming effects. The Lag10 predictor indexed the number of times participants had seen a given node in the previous 10 trials (range = 0 – 4; median = 0), while the Recency predictor indexed, for each node, the number of trials that elapsed since it was last seen (range 0 – 140; median 15). Thus, if participants viewed the sequence 1-2-5-4-3-5-3-1-3-2-5, the final node (5) would be assigned a Lag10 score of 2 and a Recency score of 5. To account for repetition priming effects, we regressed RTs from the Random and Hamiltonian conditions onto all main effects and interactions of Node Type (pre versus transition), Condition and Trial number, plus the addition of these two priming predictors, Lag10 and Recency. The repetition priming model included a random intercept for participant and by-participant random slopes for Trial, Node Type, Recency, and the interaction between Trial and Node Type (the model failed to converge with the addition of the Lag10 predictor).

**Acknowledgments**

This work was supported by NSF CAREER PHY-1554488 to DSB and an NIH grant to STS (DC-009209-12). DSB would also like to acknowledge support from the John D. and Catherine T. MacArthur Foundation, the Alfred P. Sloan Foundation, the Army Research Laboratory and the Army Research Office through contract numbers W911NF-10-2-0022 and W911NF-14-1-0679, the National Institute of Health (2-R01-DC-009209-11, 1R01HD086888-01, R01-MH107235, R01-MH107703, and R21-M MH-106799), the Office of Naval Research, and the National Science Foundation (BCS-1441502 and BCS-1631550). We are grateful to Cristina Leon for input on preliminary data analysis. The content is solely the responsibility of the authors and does not necessarily represent the official views of any of the funding agencies.

**Author Contributions**

EAK and DSB formulated the experiment. EAK, AEK, STS, and DSB designed the experiments. AEK collected the data. EAK analyzed the data. EAK wrote the paper, with input from AEK, STS, and DSB.

**Additional Information**

**Conflict of interest:** The authors declare no competing financial interests.

**Supplementary information** accompanies this paper.

NETWORK TRAVERSAL MEDIATES EXPECTATIONS                              17**References**

1. Griffiths, T. L., Kemp, C. & Tenenbaum, J. B. in *Cambridge Handb. Comput. Psychol.* (2008).

2. McClelland, J. L., Rumelhart, D. E. & McClelland, J. L. *Parallel Distributed Processing: Explorations in the Microstructure of Cognition, Volume 2: Psychological and Biological Models. Parallel Distrib. Process. Explor. Microstruct. Cogn. Vol. 1 Found.* **1,** (MIT Press, 1986).

3. Bullmore, E. & Sporns, O. Complex brain networks: graph theoretical analysis of structural and functional systems. *Nat. Rev. Neurosci.* **10,** 186–198 (2009).

4. Medaglia, J. D., Lynall, M.-E. & Bassett, D. S. Cognitive network neuroscience. *J. Cogn. Neurosci.* **27,** 1471–91 (2015).

5. Girvan, M. & Newman, M. E. J. Community structure in social and biological networks. *Proc. Natl. Acad. Sci. U. S. A.* **99,** 7821–6 (2002).

6. Cong, J. & Liu, H. Approaching human language with complex networks. *Phys. Life Rev.* **11,** 598–618 (2014).

7. Bassett, D. S., Yang, M., Wymbs, N. F. & Grafton, S. T. Learning-induced autonomy of sensorimotor systems. *Nat. Neurosci.* **18,** 744–51 (2015).

8. Goldstein, R. & Vitevitch, M. S. The influence of clustering coefficient on word-learning: how groups of similar sounding words facilitate acquisition. *Front. Psychol.* **5,** 1307 (2014).

9. Engelthaler, T. & Hills, T. T. Feature Biases in Early Word Learning: Network Distinctiveness Predicts Age of Acquisition. *Cogn. Sci.* (2016). doi:10.1111/cogs.12350

10. Schapiro, A. C., Rogers, T. T., Cordova, N. I., Turk-Browne, N. B. & Botvinick, M. M. Neural representations of events arise from temporal community structure. *Nat. Neurosci.* **16,** 486–92 (2013).

11. Fiser, J. & Aslin, R. N. Statistical learning of higher-order temporal structure from visual shape sequences. *J. Exp. Psychol. Learn. Mem. Cogn.* **28,** 458–67 (2002).

**Figures and Tables**

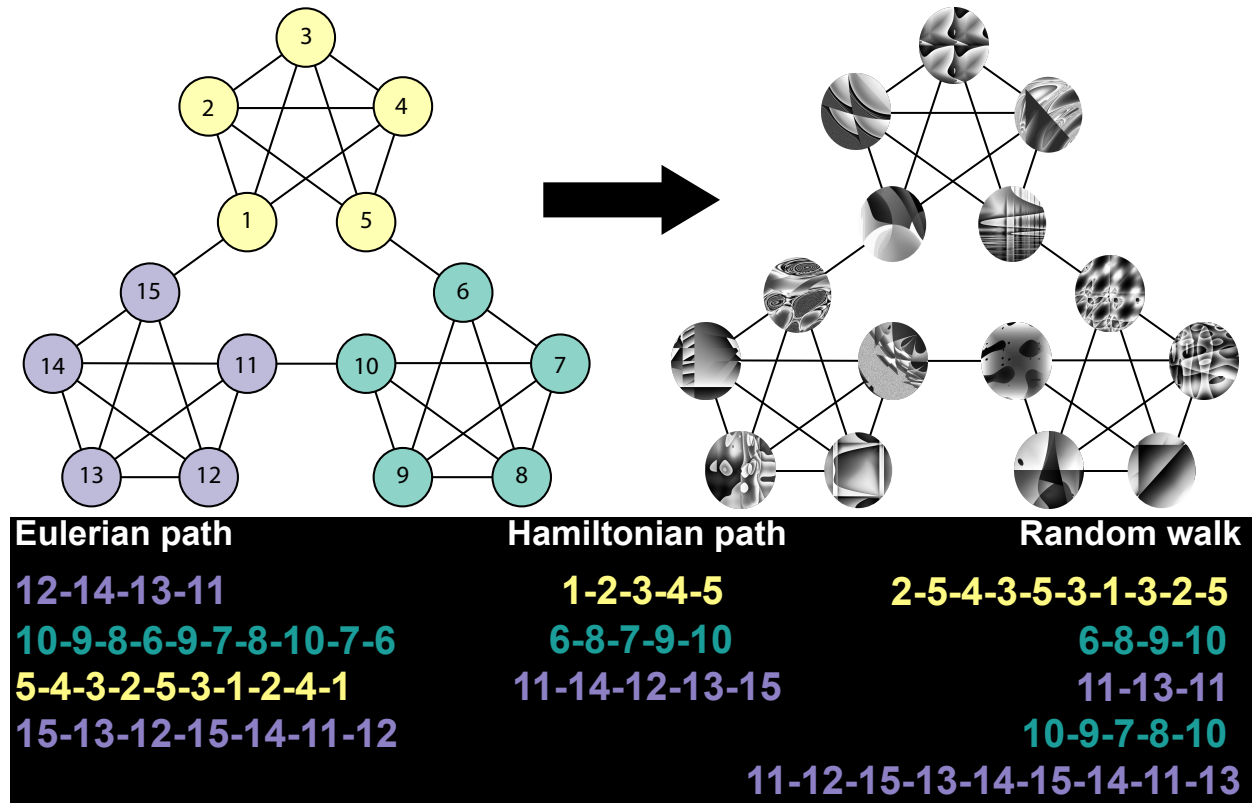

Fig. 1. Representation of the graph and walk structure underlying visual sequences. The graph consisted of three distinct communities of interconnected nodes (shown in yellow, teal, and purple). Each node in the graph corresponded to a unique fractal image, and edges between nodes corresponded to their possible co-occurrence in a sequence. Sequences were generated by "walking" along the edges of the graph randomly, or according to successive Eulerian and Hamiltonian paths. In the color-coded walk samples shown above, we illustrate that sequences generated by Random walks and Eulerian paths tended to stay within a given community (relative to Hamiltonian paths, which only sparsely sampled from each community).

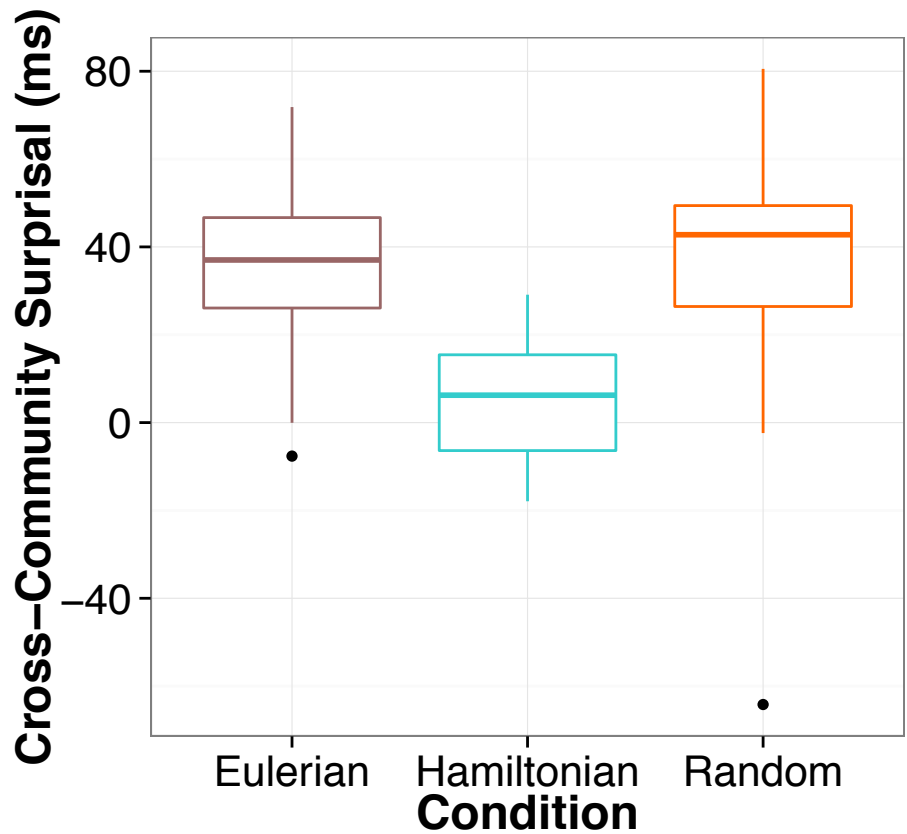

Fig. 2. Boxplots of reaction time increases across experimental conditions (N= 59). Cross-community surprisal effects were calculated by subtracting, for each participant, mean RTs for pre-transition nodes from mean RTs for transition nodes. A value greater than 0 indicates an increase in RT upon entry into a new community during the exposure phase. Note that strong evidence for surprisal is observed only for walk types involving repeated exposure to common connections within the same community (Eulerian and Random). No surprisal effect was observed for participants in the Hamiltonian condition.

Table 1. Coefficients (and corresponding t-values and p-values) for each predictor in a model examining the effect of Node Type (pre-transition versus transition), Condition (Model 1: Random versus Hamiltonian; Model 2: Random versus Eulerian), and Trial on RTs from the exposure phase. Significant values (determined using the Sattherwaite approximation and corresponding to p < 0.05) are bolded.

| Predictor | Coefficient | T-value | P-value |
| --- | --- | --- | --- |
| MODEL I | | | |
| **Node Type (pre v. transition)** | **10.85** | **6.28** | **<0.001** |
| Condition (Random v. Hamiltonian) | −5.96 | −0.55 | >0.250 |
| **Trial** | **−31.27** | **−7.72** | **<0.001** |
| **Node Type* Condition** | **−8.53** | **−4.94** | **<0.001** |
| Node Type*Trial | −2.43 | −1.54 | 0.123 |
| **Condition*Trial** | **8.28** | **2.05** | **0.048** |
| Node Type*Condition*Trial | 0.98 | 0.62 | >0.250 |
| | | | |
| MODEL II | | | |
| **Node Type (pre v. transition)** | **18.71** | **9.64** | **<0.001** |
| Condition (Random v. Eulerian) | 6.79 | 0.47 | >0.250 |
| **Trial** | **−34.17** | **−8.89** | **<0.001** |
| Node Type* Condition | −0.61 | −0.32 | >0.250 |
| Node Type*Trial | −0.85 | −0.48 | >0.250 |
| Condition*Trial | 5.55 | 1.44 | 0.157 |
| Node Type*Condition*Trial | 2.49 | 1.40 | 0.163 |

**TITLE: Process reveals structure: How a network is traversed mediates expectations about its architecture**

AUTHORS: Elisabeth A. Karuza[1*], Ari E. Kahn[2,3], Sharon L. Thompson-Schill[1,4], & Danielle S. Bassett[3,5]

AFFILIATIONS:
[1]Department of Psychology, University of Pennsylvania, Philadelphia, PA 19104 USA
[2]Department of Neuroscience, University of Pennsylvania, Philadelphia, PA 19104 USA
[3]Department of Bioengineering, University of Pennsylvania, Philadelphia, PA 19104 USA
[4]Department of Neurology, University of Pennsylvania, Philadelphia, PA 19104 USA
[5]Department of Electrical and Systems Engineering, University of Pennsylvania, Philadelphia, PA 19104 USA

*CORRESPONDING AUTHOR:
Elisabeth A. Karuza
ekaruza@sas.upenn.edu

Supplementary Information

*Section 1. Post-Exposure Measures*

**Methods**

**Materials**

**Segmentation sequence.** The segmentation sequence was identical across all three conditions. It consisted of groupings of 15 images that alternated between a random walk and one fixed Hamiltonian path. The Hamiltonian path was entered at a randomly selected node adjacent to the terminal node of the random walk, and included both backward and forward transversals. We opted to use the random/ Hamiltonian path combinations (as opposed to separate segmentation sequences paralleling each exposure sequence) in order to make a direct connection with prior work (Schapiro et al., 2013). This test format also enabled us to probe whether any learning effects from the exposure phase might generalize during exposure to a novel walk sequence. As in the exposure phase, each image was presented for 1500 ms with no interstimulus interval. The entire sequence contained 600 items.

**Odd-man out stimuli.** Items in the odd-man out task consisted of a triplet of images spanning a community boundary. Two of those images belonged to the same community (one was a boundary node), while the third was a boundary node outside that community (but sharing an edge with the other boundary node in the trio). All three images were presented simultaneously. The post-test phase consisted of 18 trials corresponding to all possible orderings of unique 3 test triplets (nodes 1, 13, 15; nodes 3, 5, 6; nodes 8, 10, 11; Fig.1).

**Procedure**

**Segmentation phase.** During the segmentation phase, participants were instructed as follows: "You have only one task: you'll see a stream of the same images presented in their regular orientation, we want you to press the spacebar at times in the sequence that you feel are natural breaking points. If you're not sure, go with your gut feeling. Try to make your responses as quickly and accurately as possible." They again completed a quiz to ensure their understanding. Participants received no performance bonus for their demonstrated ability to segment the stream. Regardless of condition they received one of two segmentation lists, counterbalanced with one of two exposure lists.

**Odd-man out phase.** In the final phase of the experiment, participants completed an odd-man out task. Trial order was randomized by subject and again, participants received no performance bonus. They were instructed as follows: "The stream of images you just saw adhered to a pattern. In other words, some of the images you saw "went together." We want to see how well you learned that pattern. For each trial, you'll be presented with three images in random order. We're interested in whether or not you can pick the single image that DOESN'T belong based on what you just saw in the previous two phases of the experiment." Trial duration was unlimited, and the start of each subsequent trial was triggered by a button press.

**Analyses**

**Data exclusions.** In the segmentation phase, we compared parsing probabilities (Fig. S1) only for those participants who surpassed the following predetermined exclusion criteria: fewer than 5 button presses per 600 trials or greater than 10 button presses in a row. These criteria resulted in exclusion of over 20% of participants (13/60). Lowered data quality on these post-exposure measures prevents us from drawing strong conclusions about the explicit expression of graph knowledge. We stress here that financial incentives based on performance, ideally on an orthogonal cover task as in the present experiment's exposure phase, are essential when collecting data in an on-line marketplace.

## Results

Participants were no more likely to segment the stream at a community boundary relative to elsewhere in the sequence (Mean difference Random: 0.029, s.d. = 0.094, $t(16) = 1.233$, $p = 0.235$; Mean difference Eulerian: 0.016, s.d. = 0.061, $t(13) = 0.990$, $p = 0.340$; Mean difference Hamiltonian: 0.024, s.d. = 0.108, $t(14) = 0.862$, $p = 0.403$). In the odd-man out judgment, participants were simultaneously presented with three images, two of which were drawn from the same community. They were asked to indicate which image "did not belong" with the others. Performance did not differ significantly from chance in any condition (Random mean = 0.267, s.d. = 0.182; versus chance, $t(19) = 1.641$, $P = 0.1173$; Eulerian mean = 0.313, s.d. = 0.184; versus chance, $t(18) = 0.486$, $P = 0.633$; Hamiltonian mean = 0.311, s.d. = 0.050; versus chance, $t(19) = 0.448$, $P = 0.659$).

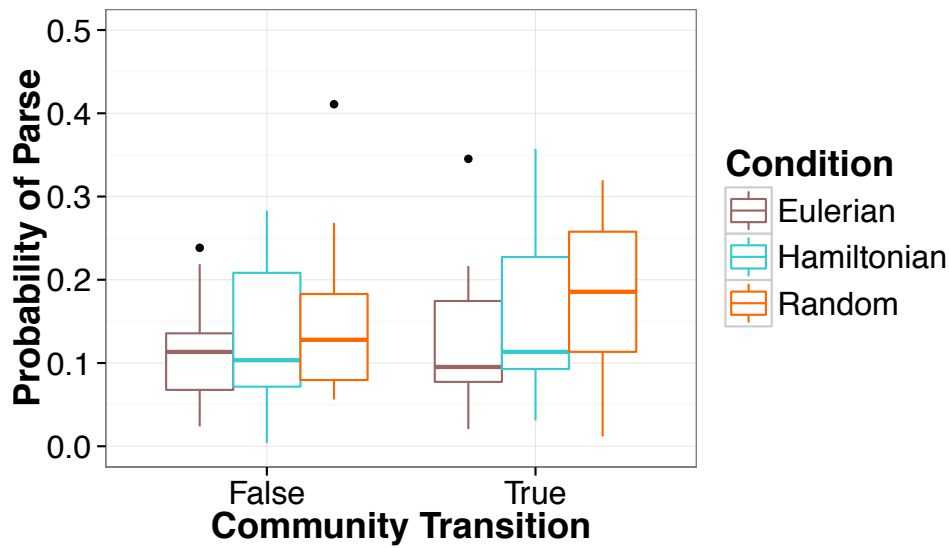

Fig. S1. Segmentation task performance. Across all experimental conditions (Eulerian, Hamiltonian, Random), participants were no more likely to parse a sequence at a community transition relative to any other node. The segmentation data presented here excludes participants who failed to comply with task directions for this portion of the experiment (remaining $N = 47$).